\documentclass[sigconf]{acmart}

\usepackage{xcolor}

\usepackage{xargs}
\usepackage{graphicx}
\usepackage[acronym]{glossaries}
\usepackage{subfig}
\usepackage{wrapfig}
\usepackage{placeins}
\usepackage{tikz}
\usepackage{array}
\usepackage{listings}
\usepackage{pifont} 
\usepackage{colortbl}
\usepackage{comment}
\usepackage{hyperref}

\newcolumntype{L}[1]{>{\raggedright\let\newline\\\arraybackslash\hspace{0pt}}m{#1}}
\newcolumntype{C}[1]{>{\centering\let\newline\\\arraybackslash\hspace{0pt}}m{#1}}
\newcolumntype{R}[1]{>{\raggedleft\let\newline\\\arraybackslash\hspace{0pt}}m{#1}}

\newcommandx{\noteReviewers}[2][1=]{\todo[inline,linecolor=blue,backgroundcolor=gray!25,bordercolor=black,#1]{\textbf{Note to Reviewers:}  #2}}

\newacronym{AM}{AM}{App Modeler}
\newacronym{CM}{CM}{Configuration Model}
\newacronym{ABM}{ABM}{App Behavior Model}
\newacronym{ABMM}{ABMM}{App Behavior Meta-Model}
\newacronym{GM}{GM}{Generator Modeler}
\newacronym{GMo}{GMo}{Generation Model}
\newacronym{AG}{AG}{App Generator}
\newacronym{NA}{NA}{Native App}

\newacronym{POI}{POI}{Point of Interest}

\newacronym{QR}{QR}{Quick Response}
\newacronym{CMS}{CMS}{Content Management System}
\newacronym{IDE}{IDE}{Integrated Development Environment}

\newacronym{MDE}{MDE}{Model-Driven Engineering}
\newacronym{MDA}{MDA}{Model-Driven Architecture}
\newacronym{BPM}{BPM}{Business Process Management}
\newacronym{BPMS}{BPMS}{Business Process Management System}
\newacronym{BPMN}{BPMN}{Business Process Model and Notation}
\newacronym{GPS}{GPS}{Global Positioning System}

\newacronym{MAML}{MAML}{M\"unster App Modeling Language}
\newacronym{DSL}{DSL}{Domain-Specific Language}

\newacronym{REST}{REST}{Representational State Transfer}
\newacronym{API}{API}{Application Programming Interface}
\newacronym{B2E}{B2E}{Business to Employee}

\newacronym{Sass}{Sass}{Syntactically Awesome Stylesheets}
\newacronym{CSS}{CSS}{Cascading Stylesheets}

\newacronym{RWD}{RWD}{Responsive Web Design}
\newacronym{HTML}{HTML}{Hypertext Markup Language}
\newacronym{JS}{JS}{JavaScript}

\newacronym{WAR}{WAR}{Web Application Archive}
\newacronym{JPA}{JPA}{Java Persistence API}
\newacronym{MVC}{MVC}{Model–View–Controller}

\newacronym{EMF}{EMF}{Eclipse Modeling Framework}
\newacronym{DOM}{DOM}{Document Object Model}

\newacronym{AmI}{AmI}{Ambient Intelligence}

\newacronym{ABR}{ABR}{Abstract Binding Repository}

\newacronym{FI}{FI}{Flow Instance}
\newacronym{BPEL}{BPEL}{Business Process Execution Language}
\newacronym{NDSPE}{NDSPE}{Native Domain-Specific Process Engine}

\newacronym{DSPML}{DSPML}{Domain-Specific Process Modelling Language}

\newcommand{\ac}[1]{\gls{#1}}

\newcommand{\acp}[1]{\Glspl{#1}}

\colorlet{punct}{red!60!black}
\definecolor{background}{HTML}{EEEEEE}
\definecolor{delim}{RGB}{20,105,176}
\colorlet{numb}{magenta!60!black}

\lstdefinelanguage{json}{
    basicstyle=\small\ttfamily,
    breaklines=true,
    literate=
     *{0}{{{\color{numb}0}}}{1}
      {1}{{{\color{numb}1}}}{1}
      {2}{{{\color{numb}2}}}{1}
      {3}{{{\color{numb}3}}}{1}
      {4}{{{\color{numb}4}}}{1}
      {5}{{{\color{numb}5}}}{1}
      {6}{{{\color{numb}6}}}{1}
      {7}{{{\color{numb}7}}}{1}
      {8}{{{\color{numb}8}}}{1}
      {9}{{{\color{numb}9}}}{1}
      {:}{{{\color{punct}{:}}}}{1}
      {,}{{{\color{punct}{,}}}}{1}
      {\{}{{{\color{delim}{\{}}}}{1}
      {\}}{{{\color{delim}{\}}}}}{1}
      {[}{{{\color{delim}{[}}}}{1}
      {]}{{{\color{delim}{]}}}}{1},
}

\copyrightyear{2022} 
\acmYear{2022} 
\setcopyright{acmlicensed}\acmConference[ICSE-SEIS'22]{Software Engineering in Society}{May 21--29, 2022}{Pittsburgh, PA, USA}
\acmBooktitle{Software Engineering in Society (ICSE-SEIS'22), May 21--29, 2022, Pittsburgh, PA, USA}
\acmPrice{15.00}
\acmDOI{10.1145/3510458.3513017}
\acmISBN{978-1-4503-9227-3/22/05}
\begin{document}

\title{Lowering Barriers to Application Development With Cloud-Native Domain-Specific Functions}

\author{José Miguel Pérez-Álvarez}
\email{jm.perez@naverlabs.com}
\affiliation{%
  \institution{Naver Labs Europe}
  \postcode{38240}
  \city{Meylan}
  \country{France}
}

\author{Adrian Mos}
\email{adrian.mos@naverlabs.com}
\affiliation{%
  \institution{Naver Labs Europe}
  \postcode{38240}
  \city{Meylan}
  \country{France}
}

\author{Benjamin V. Hanrahan}
\email{bvh10@psu.edu}
\affiliation{%
  \institution{Pennsylvania State University}
  \streetaddress{University Park}
  \city{University Park, PA 16802}
  \country{USA}
}

\author{Iyadunni J. Adenuga}
\email{ija5027@psu.edu}
\affiliation{%
  \institution{Pennsylvania State University}
  \streetaddress{University Park}
  \city{University Park, PA 16802}
  \country{USA}
}

\begin{abstract}
        Creating and maintaining a modern, heterogeneous set of client applications remains an obstacle for many businesses and individuals.
        While simple domain-specific graphical languages and libraries can empower a variety of users to create application behaviors and logic, using these languages to produce and maintain a set of heterogeneous client applications is a challenge. 
        Primarily because each client typically requires the developers to both understand and embed the domain-specific logic.
        This is because application logic must be encoded to some extent in both the server and client sides.

        In this paper, we propose an alternative approach, which allows the specification of application logic to reside solely on the cloud.
        We have built a system where reusable application components can be assembled on the cloud in different logical chains and the client is largely decoupled from this logic and is solely concerned with how data is displayed and gathered from users of the application.
        In this way, the chaining of requests and responses is done by the cloud and the client side has no knowledge of the application logic.
        This means that the experts in the domain can build these modular cloud components, arrange them in various logical chains, generate a simple user interface to test the application, and later leave it to client-side developers to customize the presentation and gathering of data types to and from the user.
        
        An additional effect of our approach is that the client side developer is able to immediately see any changes they make, while executing the logic residing on the cloud.
        This further allows more novice programmers to perform these customizations, as they do not need to `get the full application working' and are able to see the results of their code as they go, thereby lowering the obstacles to businesses and individuals to produce and maintain applications.
        Furthermore, this decoupling enables the quick generation and customization of a variety of application clients, ranging from web to mobile devices and personal assistants, while customizing one or more as needed.

\end{abstract}


\maketitle

\section*{Lay Abstract}

Creating new computer applications, distributing them to users, and maintaining them, is complex and time-consuming.
Building even a simple application requires years of study.
Furthermore, given the number of devices we have today (from smartphones to voice assistants), the task is further complicated and difficult even for experienced programmers.

The world is undergoing a digital transformation, and computer applications are now part of our everyday existence and building them is a requirement for many organizations and businesses.
However, only companies and individuals with large budgets and skills can build these applications, a consequence of which is an increased digital gap between those with means and those without.
We see a clear need to lower the barriers to building these applications so a more diverse set of voices are able to participate and shape our digital futures, as underrepresented groups are often left out of designing and building these platforms.

Previous researchers have proposed approaches where applications can be easily specified by people without (or with few) technical skills.
However, handling different device platforms and maintenance/distribution were neglected in these efforts.

In this paper we propose an approach where the program is executed on the cloud, while the client only needs to display and gather data from users.
A result of our approach is that generic front-end applications adapt to any changes to the program being executed on the cloud.
Our programs work with end-user apps on any device,  the application that the user sees automatically adapts to any changes.

\section{Introduction}
\label{sec:introduction}

Creating and customizing domain-specific applications remains an expert task, resulting in the design and development of these systems excluding voices of underrepresented groups from the design process.
This means that any groups or areas where these skills are not present, such as rural America or underrepresented groups, do not get to have a voice in determining our digital futures, and are disenfranchised to an extent.

One area of research that has tried to lower the barrier to developing applications are domain-specific approaches \cite{kelly2008domain} and \acp{DSL} \cite{van2000domain}, and the advantages of these techniques to software development \cite{mernik2005and} are clear \cite{articleDSLSurvey}.
However, there are still significant hurdles on the path between the behavioral models created by non-technical users and a fully executable application.
so, while simple domain-specific graphical languages and libraries empower a variety of users to create application behavior and logic, translating this logic into a suite of user-facing applications is out of reach for the typical domain expert.
We argue that this is in large part due to two related issues:
first, the application logic that is encoded in the behavioral models created with \acp{DSL}, needs to be represented in some way throughout each of the various layers of the application (e.g.,~the client needs to know how to interact with the behavior of the server);
second, this problem is compounded by the ubiquitous requirement for full-stack, heterogeneous clients (e.g.,~mobile, web) that each requires a different language and set of development tools.

In this paper, we present a novel approach to solving both of these closely related problems.
Our system provides methods for creating and specifying behavioral models which execute solely on the cloud.
In this system, the behavioral models are comprised of logical elements, named \textit{activities}, that are chained together to produce logical \textit{flows}.
These behavioral models encapsulate the application logic, the individual activities are reusable across behavioral models, and can be arranged in various ways (e.g.,~a login element or credit card data element can be reused).
We then isolate the behavioral model and the logic it contains from the client through our \textit{coordination mechanism}, which provides a layer of abstraction for the client, which becomes only responsible for displaying and gathering data from the user.
The behavioral model on the cloud is then responsible for persisting data and determining the next activity.
In the aforementioned login example, the client would receive a communication from the coordination mechanism that it needs to gather a text value and a password value from the user.
The coordination mechanism would prepare and format this data and send it to the cloud, where it would be connected to the current activity, and the cloud would determine whether or not the user id and password were correct and communicate the appropriate next step to the coordination mechanism (e.g., depending on whether the login was successful or not).
In this way, the coordination mechanism is what handles requests and responses to the cloud -- which is untouched by developers -- and the client has no knowledge of the application logic and is only concerned with displaying and gathering data from the user. 
This means that the client code is almost completely decoupled from the code being executed on the cloud.
Decoupling in this way enables domain experts to build these modular activities, arrange them in various logical flows, and generate a simple initial user interface for heterogeneous clients running on devices ranging from voice-based assistants to mobile phones and the web.
Later, front-end developers can customize these generated user interfaces if needed.
In this way, our system makes it easier to develop and maintain applications using \acp{DSL}.

Our method makes the development of user interfaces easier, primarily because the client-side developer is able to immediately see the effects of any code changes that they make, all while executing the production version of the behavioral model.
This effectively lowers the barrier for more novice client-side developers -- or ones who do not understand the cloud-based logic -- to do further customizations of the user interfaces. They would start with fully functional, generated versions of the user interface and iteratively change the code while seeing the results of their changes immediately.

Likewise, this approach makes the long-term maintenance of applications easier.
This is because, as domain experts push changes or corrections to logical elements on the cloud, in the vast majority of cases, consuming these updates on the client requires no technical skills or even knowledge that these updates have occurred.
Our system achieves this through this same isolation of the behavior models to the cloud, which enables these changes to be seamlessly propagated to the client side through the use of the coordination mechanism.
Contrast our approach and system with typical full-stack, multi-client applications, where each of the different front-end applications (e.g.,~iOS, Android, and web-apps) have to essentially duplicate, or mirror, much of the same application logic that has already been encoded on the cloud back-end. In such cases, any changes to the back-end must be coordinated with the front-end.
While large companies, with large software development budgets, are able to deal with this complexity, maintaining this complex software ecosystem is cost prohibitive or beyond the expertise of many small organizations and businesses.

We do acknowledge that there are other approaches for moving from domain-specific descriptions to execution, such as Mos et al. \cite{mos2016generating}, where modeled behavior was transformed into \ac{BPMN} \cite{mos2016business}. Such approaches have a somewhat tangential aim of leveraging the execution stacks of many available \ac{BPMS} in order to make it easier to deploy behavioral models, however they have the following drawbacks:
they require technical involvement from business analysts to deploy the generated \ac{BPMN} models into the various \ac{BPMS} and configure a number of custom parameters;
the behavioral models of domain-specific processes do not map directly over the various \ac{BPMN} constructs, which entails the generation of additional activities and third party components \cite{DBLP:conf/er/CornaxPML17}, thereby increasing the complexity of the resulting artifacts;
they introduce a dependency to an additional language, and more importantly, to a full execution stack that needs to be acquired and managed separately;
they do not have mechanisms to isolate client-side code from changes on the server-side, which complicates maintenance activities.
Therefore such approaches do not help achieve the specific goals of this proposal, which are centered around making it easier to develop, deploy, and maintain full-stack applications through the decoupling of the front-end from the back-end code.

We evaluate our approach in two main ways.
First, we demonstrate the flexibility and validity of the method by implementing the \textit{Conduit} application, which is a clone of \textit{medium.com}, as described in the \textit{RealWorld} project \cite{Sun2019}.
We use this large open-source project as a programming benchmark, as there are many competing implementations against which we can compare our approach.
Second, we validate that novice-level programmers (i.e.~students who have taken an introductory programming course) are able to customize a simple generated client app without being aware of the application logic on the cloud.

The paper is organized as follows:
Section \ref{sec:overview} introduces the overall architecture and the concepts of our proposed system, where we outline how the behavioral models are defined and executed, and how the coordination mechanism enables the decoupling of the client and cloud components.
Section \ref{sec:evaluation} describes the two evaluations that we performed on our system:
first, we describe a prototype we developed using the \textit{Conduit} scenario to compare our implementation with other languages and technology stacks; 
second, we carried out a user evaluation with novice developers who performed different customizations of a simple system.
Finally, Section \ref{sec:related} presents the most relevant related work and contrasts it with our approach, and Section \ref{sec:conclusions} provides our final thoughts and future plans for our system.

\section{System Overview}
\label{sec:overview}

\begin{figure*}[htb]
	\centering
	\includegraphics[width=\textwidth]{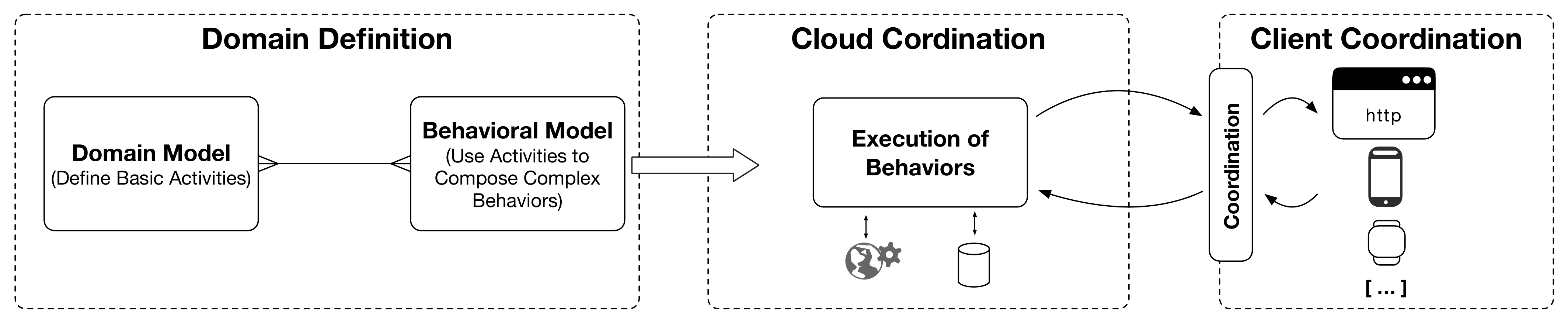}
	\caption{System overview: the different components of our system are grouped into the Domain Definition, Cloud Execution, and Client Coordination. Each group has different responsibilities and is essential to the decoupling of cloud and client.}
	\label{fig:elements_involves}
\end{figure*} 

In this section, we outline our system and how we have designed it to decouple the behavioral models and user applications, effectively decoupling the cloud-side logic from the client front-ends.
The key to how we achieve this decoupling is our \textit{coordination mechanism}, which provides a set of abstract instructions that reduce or eliminate the responsibilities of the client code in communicating with the cloud. 
The cloud components of our system are responsible for isolating data between different applications, locations, and users.
These components are additionally responsible for preserving state continuity of individual user interactions and persisting data of and between sessions.
This continuity is achieved without the client code needing to know or be concerned with how data is being managed and carried forward to future interactions by our system.
This isolation means the client code in our system does not depend on the cloud application logic, and enables the system to generate generic end-user interfaces which can later be easily customized.
Building our system in this way, allows the rapid bootstrapping of heterogeneous client applications with progressively unique user experiences.

Figure \ref{fig:elements_involves} illustrates the different elements contributing to the approach our system embodies.
There are three distinct components in our system that we describe in this section:
the \textit{Domain Definition} component, where technically skilled domain experts create the \textit{Domain Model} by defining the basic activities supported by the domain (e.g., login, retrieve data) and the \textit{Behavioral Model} that assembles these activities into meaningful logical flows (e.g., buying an item or viewing users);
the \textit{Cloud Execution}, where deployed behavioral models are executing on the cloud and interacting with databases and external services as needed;
and the \textit{Client Coordination}, where clients are ushered through the different activities in the behavioral model and instructed to either display or gather data from the user.
The focus of this paper is the \textit{Cloud Execution} and the \textit{Client Coordination}, particularly in how the coordination mechanism enables client customization and seamless pushing of changes to the behavioral models.

\subsection{Domain Definition}

While a more detailed overview of these three modeling elements are out of the scope of this paper (they are available in our previous work \cite{arXivPaper}), some amount of background is needed for the \textit{Domain Definition} where domain experts specify the \textit{Domain Model} and \textit{Behavioral Model}.
In our system, users with technical knowledge create individual \textit{Activities} as part of different \textit{Domain Models}, these domain models are essentially a library of available activities that are logically grouped around a domain, for example a shopping domain would include payments or object details, etc.
These activities are reusable and support many different types of functionality, such as connecting to a database, defining inputs/outputs, connecting to external web services, user login, etc.

Using these activities, users with less technical knowledge then create \textit{Behavioral Models} where they assemble these activities and different non-functional elements (e.g.,~starting/ending points, and decision/looping constructs) into logical flows, such as buying an item or retrieving articles from a site.

\begin{figure}[ht]
	\centering
	\includegraphics[width=0.52\textwidth]{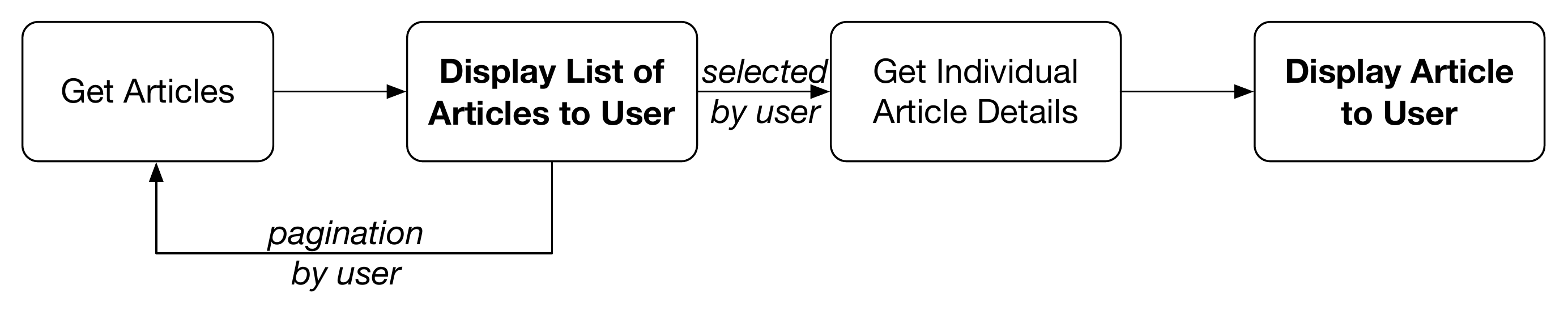}
	\caption{Example behavioral model: while all activities execute on the cloud, only activities in bold are communicated to the client through the coordination mechanism.}
	\label{fig:proces_model}
\end{figure}

A sample flow that models the behavior for retrieving the list of articles in our \textit{Conduit} evaluation scenario, can be seen in Figure \ref{fig:proces_model}, where each rectangle represents an activity and each arrow a transition element. 
The first activity, \textit{Get Articles}, is defined to retrieve a set of articles from a database (or alternatively an external service).
Once these articles are retrieved from the database the logical flow indicates that the article list will be displayed to the user. The domain expert specifies aspects of this activity such as the article data type, what needs to be displayed to the user, and what information needs to be gathered from the user.
In this example, the data gathered from the user is if they would like to request more articles (which activates the pagination transition) or select an individual article to view (which activates the selection transition).
If the user selects an individual article, the behavioral model specifies that the next activity will be to retrieve the details about the individual article from the database and proceed to the the final activity, where the behavioral model specifies to send the individual article to the client to display to the user.
Throughout the process of assembling activities into these logical flows, domain experts can experiment and interact with them through the generated clients, without writing any client code.
In the following section, we detail how these behavioral models are executed on the cloud.

\subsection{Cloud Execution}

\begin{figure*}[t]
	\centering
	\includegraphics[width=\textwidth]{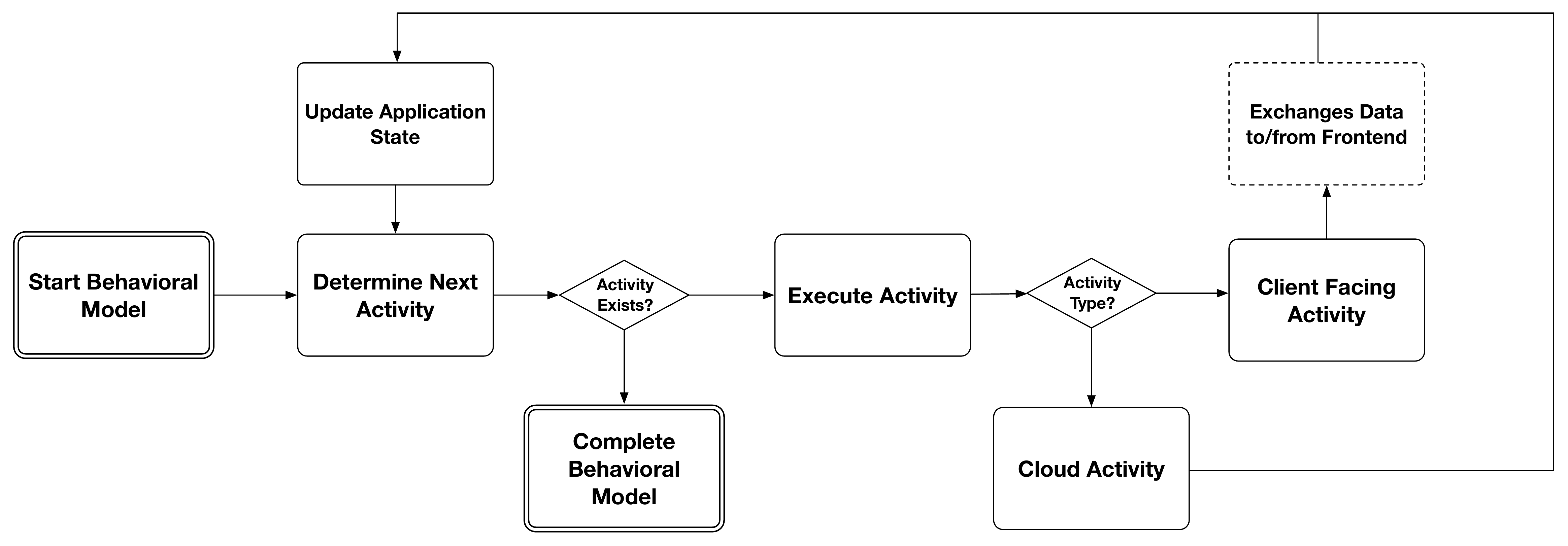}
	\caption{Behavioral model execution. The focus of this paper is how the client and cloud communicate throughout steps \textit{Client Facing Activity, Exchanges Data to/from Frontend, and Update Application State}.}
	\label{fig:Idea}
\end{figure*}

The behavioral flows are executed solely on the cloud, effectively isolating the application logic from the client which only needs to display and gather data from the user.
Figure \ref{fig:Idea} gives an overview of this execution and details when the system exchanges data to/from the client.

In the first step, we create an instance of the behavioral model, in this instance we store information about the application state.
This application state data is in many ways similar to session data in a web application, however, it differs in that it contains additional information specific to our system (e.g.,~what the current activity is, global variables for the entire model, and other system status/health information).

Once the instance has been created, the execution engine running on the cloud enters the main execution loop of the behavioral model. This is a continual loop where it determines which activity needs to be executed and then proceeds to start the execution of that activity.
During the execution of the activity, the engine uses the activity metadata to determine if it is a user facing operation or not.
This is defined in the activity by whether or not it requires displaying or gathering data from the user, if user interaction of some sort is required, then the cloud engages with the coordination mechanism (part of which runs on the cloud and part of which runs on the client, indicated by the dotted outline).
Whatever the nature of the activity being executed, upon completion, the application state is updated and the loop continues.
Once the model has reached the final activity, the execution terminates.

Additionally, there are a few details on cloud execution worthy of note.
We do support parallelism, in that there can be multiple starting points in the behavioral model and our system queues the different activities, if there are dependencies between the activities our system resolves them.
Each execution of an activity also creates an activity instance, where specific details and data needed to complete that activity are stored.
Once the activity has finished executing, these instances are destroyed.

\subsection{Coordination Mechanism}
\label{sec:integration}

This section focuses on the main contribution of this paper which details how we automatically execute application logic residing on the cloud and how it integrates with the client application.
This automatic integration enables the client code to be completely decoupled from the logic on the cloud.
Decoupling in this way allows clients to be written in a very generic way.

\begin{figure*}[t]
	\centering
	\includegraphics[width=0.95\textwidth]{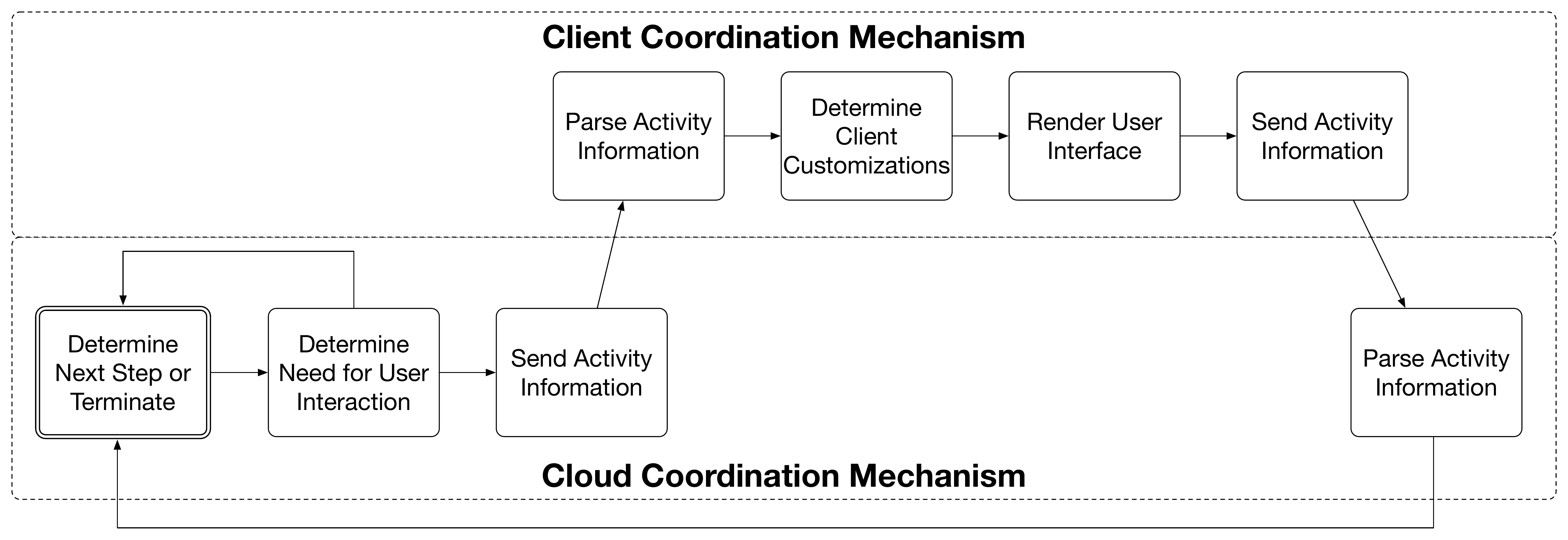}
	\caption{Communication Model}
	\label{fig:communication_model}
\end{figure*}

Figure \ref{fig:communication_model} illustrates how the coordination mechanism orchestrates the execution of the behavioral model between the cloud and the client.
When the client first launches a behavioral model through the front-end, the engine retrieves the behavioral model specification and starts executing it, entering the main loop.
The cloud coordinator executes each subsequent activity on the cloud, and continues execution until a client facing activity is encountered.
Once a client facing activity is encountered, the cloud coordinator serializes the application and activity state information that the client coordinator will need to proceed.
Essentially, this information is a JSON dictionary that outlines the elements to display to the user, the elements to gather from the user, any constraints that the client coordinator must satisfy before sending a response, and any data objects to display.
This dictionary does not contain any information about how to interact with the end user (no HTML, CSS, or other user interface code). 
This is because the cloud coordinator completely delegates the responsibility of interacting with the end user to the client coordinator, and it will depend on the characteristics of the client device (e.g., whether it has a screen or not, the available sensors). 
For example the data that would be sent in Figure \ref{fig:proces_model} to display articles to a user would be:

\begin{lstlisting}[language=json]
{"instanceId": 15,
 $\textbf{"displayElements"}$: [
  {"name":"alist", "label":"Article List",  
   "set":true, 
   "subElements": [
    {"name":"title", "label":"Article title",
     "type":"String"},
     {...},]}],
 $\textbf{"gatherElements"}$: [
  {"name":"selected", "label":"Article selected",
   "set":false, "type":"String"},
  {"name":"pagination", "label":"Get more",
   "set":false, "type":"boolean"}]
 $\textbf{"constraints"}$: [
  {"name":"selected", "valueFrom": "alist"}]] 
 $\textbf{"value"}$: [
  {"alist": [{"id":1, "title":"How to create the behavior model"}, {...}]}]} 
\end{lstlisting}

When the client coordinator receives this information, it is able to determine the basic aspects of how to render the user interface in order to both display and gather information from the user.
\begin{itemize}
    \item The \textit{displayElements} portion tells the client coordinator what elements to display, along with a label, and in this case information describing the structure of the individual elements of this set.
    \item The \textit{gatherElements} portion tells the client coordinator what elements to gather from the user, in this case there are two options, selecting an individual article or to request more articles.
    \item The \textit{constrains} portion tells the client coordinator whether or not there are any further requirements for the elements to gather from the user, in this case the 'selected' element is a value from the value 'alist'.
    \item Finally, the \textit{value} portion tells the client the actual values to display to the user as described in displayElements.
\end{itemize}

Once the user has satisfied these conditions using the user interface, the client coordinator is able to send a message back to the cloud coordinator about the result of the interaction.

\begin{lstlisting}[language=json,firstnumber=1]
{"instanceId": 15,
 $\textbf{"response"}$: [
  {"selected":{"id":1, "title":"How to create the behavior model"}}}
\end{lstlisting}

The cloud coordinator receives this information and determines the next step, in this case to get the individual article from the database and send it to the client coordinator.
What is worth noting here is that the cloud coordinator is able to keep track of which activity instance a particular \textit{http} response is in regards to with the \textit{instanceId}, as such the cloud engine is able to juggle many parallel instances of activities and behavior models.
Furthermore, the client coordinator is not aware of the overall execution of the behavior model, and only knows the current activity, what data are provided by the cloud coordinator, and what data is needed by the cloud coordinator. 

This is why changes to the behavior logic on the cloud do not affect the clients. Consider an example: in the \textit{Conduit} scenario, the list of articles does not contain a representative image. If the domain designers wish to correct this, they can add it to the activity definition in the \textit{domain definition} on the cloud-side. This will in turn entail an automatic update to the information sent by the cloud coordinator which will become:

\begin{lstlisting}[language=json]
{"instanceId": 15,
  "displayElements": [
  {"name":"alist", ..., 
   "subElements": [
    {"name":"title","label":"..","type":"String"},
    {$\textbf{"name":"rimage",}$ $\textbf{"label":"Representative image",}$ $\textbf{"type":"Image"}$}
     {...},]}],
 "gatherElements": [...]
 "constraints": [...] 
 "value": [
  {"alist": [{"id":1, "title":"How to create the behavior model", $\textbf{"rimage":"http://bit.ly/1"}$}, {...}]}]} 
\end{lstlisting}
However, the client code does not change at all and need not be aware of the change on the cloud side. The client coordinator will simply manage the updated user interface based on this information.

Although it is out of the scope of this document, it is also possible to include extra information in the exchanges, such as a customer identifier.
This could be used by the engine, for example, to disambiguate the origin of the data in case of multi-client execution of the same task, or indeed to qualify certain properties of sensor data depending on the device.

One important consequence of decoupling the logic on the cloud, is that clients for any target device and platform, being highly generic with respect to application logic, can be systematically generated to run natively on the user platform. \cite{DBLP:conf/icse/Perez-AlvarezM20}.
These native clients can therefore take advantage of the specific features and sensors on that platform, and use them in the application logic.

\subsubsection{Enabling Client Side Customizations}

In this section, we describe how the combination of components and abstractions in our system enables easier development and customization of heterogeneous client applications.
The majority of these customizations are brought in during the \textit{Determine Client Customizations} stage in Figure \ref{fig:communication_model}.
Here the coordination mechanism makes a series of checks for client customizations.

First, the coordination mechanism checks if there is a specific custom implementation for presenting the current activity.
This is the highest form of customization in our system, here the front-end developer can create their own bespoke layout that applies \textit{only} to that particular activity as a whole.
In a web front-end for example, this is an HTML file named with an identifier that includes the same name as the activity.
The communication of this layout between the client and the cloud is still handled entirely by the coordination mechanism, however, the rendering of the user interface is defined by the front-end developer.

The second type of customization that the coordination mechanism checks for -- if there is no specific customization for the activity -- is represented by customization layouts for the different variable types.
Here the rendering of the user interface is still handled almost entirely by the coordination mechanism, however, when the mechanism renders a specific type of field (e.g., a boolean), it will check to see if the front-end developer has provided an overriding implementation.
That is, as the client coordinator is looping through the different display and gather elements, it renders a different snippet (e.g., of HTML) in that slot.
If the client developer has created an HTML template to override that snippet, that custom template is put into the slot in the user interface.

\subsubsection{Seamlessly Propagating Cloud Side Changes}

The same mechanisms that enable easier client application development, also make it easier to propagate changes from the cloud to the client.

The vast majority of changes require no additional development in the client application. For instance, changing the label of an attribute related to an activity of a domain would automatically be taken into account when rendering the appropriate input and output data handling elements in the user interface. In addition, adding any number of attributes to the data used in activities, provided these types are supported by the client coordination code, would also not involve any change in the front-end code. Even more importantly, changing the logic of the behavior models, including reordering existing activities, updating conditions on the transitions between activities, adding any number of new activities at any place in the behavior models, would not require any changes in the client application. The only times changes are needed on the client is when customizations are needed to support very specific activities or indeed to support new primitive data types used in activities (e.g., video or 3D images if not already supported). 

This effectively means that the overall coordination mechanism nearly ensures that any client application that is currently functioning will continue to function regardless of changes to behavioral models on the cloud. The coordination mechanism in the client applications with automatically adapt to the new behavioural model.

\section{Evaluation}
\label{sec:evaluation}
To evaluate the efficacy and validity of the approach proposed in this paper we performed two evaluations.
First, we needed to validate that our system was capable of implementing a realistic application, and compare the amount of effort involved to other implementations.
To do this we chose the \textit{RealWorld} \cite{Sun2019} \textit{Conduit} benchmark, for which there are multiple implementations.
Second, we needed to validate that a novice would be able to make progressively complex modifications to the generated user interface, for this we did an evaluation with undergraduate students at the start of their programming sequence of classes.
Both of these evaluations are presented below.

\subsection{Prototype Evaluation}
\label{sec:prototype}

In order to evaluate the efficacy of our system, we developed a fully operational prototype of the system, which provided the functionality needed to design and execute behavioral models, as well as to communicate with client code through the coordination mechanism.
We did this using several technologies.
The main technology that we used for the front-end of the prototype (a web portal for designing and managing models) is the open source framework \textbf{Angular} \cite{kasagoni2017building}, maintained by Google.
On the server, for the cloud-based engine, we used \textbf{Spring} \cite{gutierrez2016pro}, which provides components for many purposes such as inversion control, managing web requests, security or data access.
Our prototype follows the \ac{MDE} methodology \cite{mde}, with \textbf{\ac{EMF}} \cite{Steinberg:2009:EEM:1197540}, as well as \textbf{Xtext} \cite{bettini2016implementing}, as the main technologies for managing the different models.
We use the Tamgu language \cite{tamgu} for rich definition of conditions and expressions in behavioral model flows. 

We used our prototyped system to implement the \textit{RealWorld} \cite{Sun2019} \textit{Conduit} multi-technology benchmark.
Since \textit{Conduit} is a blogging application, its main element is an \textit{article}, containing a \textit{title}, \textit{subtitle}, \textit{body}, set of \textit{tags} related to the content, as well as meta-data.
\textit{Conduit} offers functionalities that allow the user to write new articles, retrieve existing articles, remove articles, add comments to articles, among others.
An important advantage of using \textit{Conduit} as a benchmark is that it has a very clear separation between client-side and server-side, with precise specifications of what can be placed on each. All the different versions that implement it follow the same strict separation.

Our evaluation strategy was as follows.
First, we created a full-blown clone of the \textit{Conduit} \textbf{cloud-side} functionality using our prototype by specifying the application logic for its various scenarios.
Second, we created two fully-fledged \textbf{clients} that behave identically to the \textit{Conduit} clients, by leveraging the coordination mechanism for two distinct platforms.
Third, we measured the effort that was required to implement three existing versions of the original \textit{Conduit} benchmark, and compared it to the effort required to implement the same functionality using our approach.
Fourth, we proposed a meaningful change in the application logic and we implemented it by modifying the client and cloud code bases of the same three \textit{Conduit} versions as well as of our version.
Lastly, we measured and compared the impact of this change across these code bases.

\subsubsection{Our Implementation of \textit{Conduit}}

Our approach implies the specification of cloud-side reusable logic as models, so it doesn't require users to write code.
Naturally, external services need to be implemented and made available but this is the case for any approach, and the \textit{Conduit} benchmark provides these services for all its implementations.
Our approach enables the declarative, wizard-based configuration of the reusable units of behavior (activities) that can leverage external services.
The only additional service activity beyond our basic functionality that our implementation needed was for processing Markdown text, and we implemented it using available libraries, in about 10 lines of code.
The rest of the service activities that other \textit{Conduit} versions use, were supported directly by built-in functionality available in our engine.

To develop the models we created one supporting domain definition with 20 activities, 9 services, and 3 different data Types.
This task took one full day, we acknowledge however that as the creators of the system, we are far more familiar with the system than real users would be, however, this does give a measure of the effort involved.
Implementing the application logic required the creation of 13 different logical flows.
This task took another day, so two total days of work to implement the cloud side.

Finally, we wrote two clients that use the coordination mechanism to access the cloud-based application logic.
One of the clients, with almost identical look and feel to the ones of the \textit{Conduit} benchmark versions, is implemented as a mobile application for iOS by using the \textbf{Ionic Framework} \cite{yusuf2016ionic}.
The other client is implemented as an Amazon \textbf{Alexa skill} \cite{Alexa:2017:AAU:3137257} so its user interaction is performed via speech recognition and generation, all while executing the same application logic.

In order to facilitate debugging the server and client code, our prototype offers a set of capabilities to the user.
First, while developing flows, the flows are in a special state called \textit{sandbox}, where the user can make modifications without impacting production versions.
This enables the user to observe what is happening in the executing flows, as well as those that have finished. 
Users are also able to modify the data received by the different flow instances that are executing, and view the impacts of these changes.
The client applications do not have a special set of tools, and therefore are debugged with the standard debugging tools of the platform.

\subsubsection{Comparative Evaluation}

In order to provide a meaningful evaluation, we systematically extracted the relevant code for the same functionalities, from several versions of the \textit{Conduit} benchmark, using its open-source \textit{GitHub} repository \cite{realWgb}.
We counted the number of files (excluding certain non-essential ones) and the lines of code (LOC) across these versions for the server code as well as the client code.
We did the same for our approach.
Since our approach involves mostly models (on the cloud side), we generated textual version of the graphical models using automated tools, and counted those lines as well.
The reasons for comparing lines of actual code with lines of model descriptions is that the alternative would be to compare lines of code with graphical constructs.
We believe that any creation (code or models) requires some expertise and it would be misleading to say that models, because they are graphical, are equivalent to zero code.
This is why we use this approach, which, while not perfect, gives a good indication of complexity, size and ultimately human effort required to create them.

In the previous subsection, we have provided the effort of our implementation of \textit{Conduit} in days. 
However, since we do not know the number of hours invested in implementing the different versions of the \textit{Conduit} benchmark, we believe that LOC can provide a proxy for the level of complexity of the different implementations.

The tables in this section highlight the differences between the ``classic'' approach used in three different versions of the \textit{Conduit} benchmark, and our approach.
The tables describe the \textbf{baseline} (initial code-base with the \textit{Conduit} standard functionality) in columns \textit{B.~Files} and \textit{B.~LOC} as well as a modified version (for which we report the total number of modified files in \textit{M. Files} and the strict \textbf{delta}, the new code and files with \textit{{$\Delta$ LOC}} and \textit{$\Delta$ Files} respectively).
The modified version is simply an updated version that adds support for adding an image to articles.
This is a simple, reasonable change, that one could imagine such an application would need. 
It involves changes on the cloud side and it must also be visible on the client side.

Table \ref{table-server} focuses on the cloud-side application logic, and Table \ref{table-client} on the client-side.
While the contribution of this paper is mostly concentrated on the cloud-side representation of behavior as reusable functions, it is important to highlight the positive impact of the coordination mechanism on the clients.
This is visible in a dynamic, change management scenario, such as the one described above.
For the client-side comparison we used only one \textit{Conduit} version because no further insight could be extracted from more versions.

\begin{table}
\begin{tabular}{|c|c|c|c|c|c|}
	\hline 
	& \textbf{B. Files} & \textbf{B. LOC } & \textbf{M Files} & \textbf{$\Delta$ Files} & \textbf{$\Delta$ LOC}  \\ 
	\hline 
	\textbf{Django} & 38  & 1064 & 10 & +6 & +31 \\ 
	\hline 
	\textbf{Express} & 14 & 2442 & 7 & +2 & +36 \\ 
	\hline 
	\textbf{Spring} & 103 & 4124 & 12 & +6 & +194 \\ 
	\hline 
	\textbf{Ours} & 15 & 560 & 1 & 0 & +5  \\ 
	\hline 
\end{tabular} 
\caption{Cloud Code}
\label{table-server}
\end{table}

The most striking difference between the Conduit versions and our approach in Table \ref{table-server} is visible in the baseline implementations.
Certainly, writing thousands of LOC for the baseline versions, even in the most optimistic estimates, spans at a minimum, several weeks of development, significantly longer than with our approach (as mentioned, the LOC for our approach include mostly the textual representation of the graphical models for domain, behavioral model flows and bindings to external services).
This is partly explainable by the built-in support for flow execution that includes robust data management capabilities, both for transient data-flow and external persistence.
When changes to existing application logic are required, the advantages of our approach (as seen in the $\Delta$ columns) revolve around easy re-composition of the flows and one-shot application of domain changes across multiple flows.
For instance, adding support for the article image involves lightly modifying the domain, but doesn't require any change to existing flows.

\begin{table}
\begin{tabular}{|c|c|c|c|c|c|}
	\hline 
	& \textbf{B. Files} & \textbf{B. LOC }  & \textbf{M Files} & \textbf{$\Delta$ Files} & \textbf{$\Delta$ LOC}  \\ 
	\hline 
	\textbf{Angular} & 105 & 2631 & 15 & +1 & + 106 \\ 
	\hline 
	\textbf{Our iOS} &  62 & 2488 & 0 & 0 & 0 \\ 
	\hline 
	\textbf{Our Alexa} & 19 & 784 & 0 & 0 & 0 \\ 
	\hline 
\end{tabular} 
\caption{Client Code}
\label{table-client}
\end{table}

Table \ref{table-client} gives an overview of the effort required to implement fully functional \textbf{generic} client applications for our approach using different technologies (and contrast them to a state-of-the art client implementation specific to \textit{Conduit}).
Particularly, this table highlights the impact of change on these clients.
It is notable that while these are fully functioning client applications on very different platforms, they need no change to support the update in application logic.
While on a mobile phone article images are displayed as expected, on the voice platform, image support takes different forms.
In the simplest version (which we took for this paper), the agent speaks out the image label.

\subsection{User Evaluation}
\label{sec:user_evaluation}

In our user evaluation we evaluated whether novice programmers would be able to effectively utilize our cloud functionality, while customizing the corresponding generated user interfaces.
In line with this goal, we recruited six (5 males, 1 female) participants from the second programming class in the College of Information, Sciences and Technology at a 4-year college to evaluate a simple scenario.
Participant ages ranged from 19-21 and included all class levels (freshman, sophomore, junior and senior).
We recruited our participants at the start of the semester, as such, they had only had an introductory programming course at the time of this evaluation.

\begin{figure}
\centering
  \frame{\includegraphics[width=0.9\columnwidth]{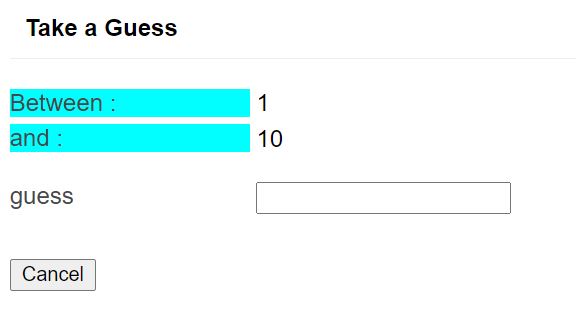}}
  \caption{Take A Guess template}~\label{fig:takeAGuess}
\end{figure}

\begin{figure}
\centering
  \frame{\includegraphics[width=0.9\columnwidth]{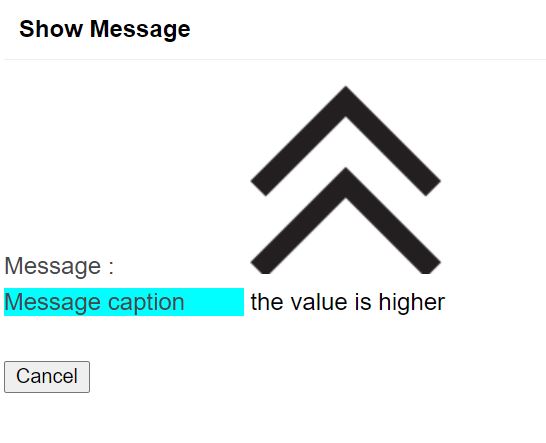}}
  \caption{Show Message template}~\label{fig:showMessage}
\end{figure}

In order to evaluate our approach, we created a guessing game web application and asked participants to customize the user interface of a web-client.
This application had two connecting steps (templates) called \textit{Take-a-Guess} and \textit{Show-Message}.
The Take-a-Guess template (Figure \ref{fig:takeAGuess}) asks a user for their guess while the Show-Message template (Figure \ref{fig:showMessage}) informs the users of the correctness of their guess.
The default HTML templates for these steps were generated and provided to participants.
The interviews were conducted via Zoom and the users performed most of the tasks described below with the HTML templates on a JetBrains WebStorm IDE while viewing the results on a web page. 

There were 3 parts to this evaluation.
There was a pre-interview where users were asked intermediate level programming questions to gauge the participants programming experience, e.g.~server architecture and JSON objects.
We also asked about their general knowledge of HTML/CSS and whether they had ever created a web application.
We then gave them a brief overview of our system (where snippets are defined) before they completed the tasks described below.

\subsubsection*{Tasks and Procedure}
We asked the participants to complete three tasks. 
In the first task, the participants had to change the background color of the texts in the default Take-a-Guess template.
In the second task, the participants had to replace the congratulations image shown to the user when they enter the correct guess.
The third task was the most complex, participants needed to create their own custom Take-a-Guess template. 
We gave the participants minimal details, in that we pointed out the values that were available (the details are rendered in the user interface as one customizes the template).

After attempting to complete the tasks, the participants were asked questions to understand their experience with the system and questions to see their knowledge of the cloud architecture.
The participants that took part in this evaluation had no web application experiences and 5/6 had no experience with HTML or CSS. 
The one participant who did have experience with HTML, had taken a first year HTML class.
All participants had beginner programming knowledge in either Python or Java, no participants had knowledge of cloud architecture (e.g.~how data is communicated between the cloud and the client).  
The responses from the participants, labeled P1 - P6, are discussed below.

\subsubsection{Levels of User Customization}

The difficulty of the tasks increased as the evaluation progressed. 
Even though the participants were able to complete the variety of tasks from simple modification of existing templates to creation of their own templates, they mentioned that they found modifying the default templates by changing the colors, images very much easier than having to design their own templates which is as expected.

The different types of Take-A-Guess templates created by the participants were mainly concerned with changing the manner in which the text was displayed, however, some participants attempted different designs and modified the position of the text and different validations on the user input. 

\begin{quote}
    Just design aesthetic wise, I would probably like that.  Take a guess part would be better, if it was centered and more bold, I would like that.- P6
\end{quote}

\subsubsection{Understanding the Interface Design Tool}

The participants had to customize and design the interfaces with HTML. Due to the absence of experience in HTML, it was initially difficult for the participants to create their template customizations.
These comments show the lack of experience that our participants had with HTML, and they were still able to complete the tasks. 
\begin{quote}
    because I mean code can be  confusing when you first look at it and you don't  know exactly what it is you're looking at - P2
\end{quote}

\begin{quote}
    There [were] some parts where I was really confused on [for example] the labels [...] and like syntax structures. - P6
\end{quote}

We gave the participants a very brief explanation of some HTML tags if they asked, but primarily they found that interacting with these tags for some time was informative -- especially as they could see the results within the application in real-time.
\begin{quote}
    But once I got used to them, [it was] fairly simple to customize the web page - P5
\end{quote}

\subsubsection{Understanding the Logic}

Based on their comments, the participants did seem to gain high level knowledge of the cloud-side logic employed by the system through interacting and designing an interface for the system.
They basically defined the steps and conditions used in the creation of the flow for the game.

\begin{quote}
     [The system] matches it to see if it is equal to the stored data, which is already five to see if it is correct. - P1
\end{quote}

\begin{quote}
    Oh, so maybe just takes the number and then goes through like 
an algorithm and sees if it's higher or lower - P6
\end{quote}

\begin{quote}
    Supported as like as an input and then the program compares it to see whether it's bigger or smaller than the actual number and then if it's bigger, it tells you to pick a lower number, and if it's smaller tells you to pick a higher number - P3
\end{quote}

While the behavioral model that we used for this evaluation is simple, each step is representative of the level of complexity in a larger, more complex behavioral model.
This is because, when you isolate the application logic from the client, the client developer is only concerned with modifying the rendering of an individual step, in the same manner in which the participants did.

\subsubsection{Ease of Use}

In terms of learning how to modify code and see how it impacted the user interface, this enabled our participants to quickly understand how to make their modifications.
Keep in mind that many of our participants only have experience with the Java language, and had never worked with HTML or CSS before.

\begin{quote}
    In the span of 20 minutes, the fact that I was able to learn that much through trial and error it was pretty impressive. - P1
\end{quote}

Isolating the individual activities and being able to use already existing, running code proved important for these novice developers.
Not having to worry about getting the code running before being able to see changes, helped them to feel that it was a more tractable problem.

\begin{quote}
    That the code is it's there and it's not daunting like some code can be sometimes it's a small exercise for just examples you can kind of look at and with very little knowledge. - P5
\end{quote}

\begin{quote}
    I like that it had like immediate feedback and I could see what I was doing and see the changes immediately - P3
\end{quote}

Some participants wanted a more cohesive environment with what-you-see-is-what-you-get operations, which can be provided with different IDEs that are more aimed at novices.
This could be included in future more feature complete versions, or they could use other development environments that provided this capability.

\begin{quote}
I was expecting [to be able to change] the website itself, [...] like drag and drop stuff  - P6
\end{quote}

All the participants with just little programming knowledge created multiple functional and seamless versions of the guessing game without any change in the cloud logic.
Even though the interface changes were mainly textual, the versions were unique to each user and some participants expressed interest in wanting to augment the interface designs using images, etc.
Also, while the participants gained the general logic of the guessing game from creating their interface, they did not need to learn the technical aspects of the logic being executed on the cloud to achieve their goals.

\section{Related Work}
\label{sec:related}

We consider our contributions in this paper structured around three main axes.
The first relates to the \textbf{model-based description of reusable behavior}, which, combined with the \textbf{coordination mechanism}, completely decouples client interaction from the logic being executed on the cloud.
The second relates to the \textbf{execution and sharing of the application logic on the cloud}, enabling support for heterogeneous client platforms, seamless updates and management, among others.
The third relates to the possibility of having \textbf{generic, easily customizable clients}, which allows having applications that work with any logic the cloud, as well as being customizable by users without much technical knowledge.

Related to the first axis, significant amount of work targets the definition of behavior as models or high-level specifications.
Such behavior, seen as higher-level than programming code, and targeting less technical roles, is also referred to as \textit{business logic}, typically defined using \acp{DSL} or graphical languages.
\acp{DSL} are typically used to improve the productivity of developers and the communication with domain experts \cite{fowler2010domain}.
As such, they are still often relatively low-level and in reality rather technology-oriented.
For instance, many \acp{DSL} target specific concerns in application development.
YANG \cite{bjorklund2016yang} is a data modeling language used to model configuration data, state data, Remote Procedure Calls, and notifications for network management protocols.
ADICO-Solidity \ac{DSL} \cite{frantz2016institutions} is designed to allow the specification and interpretation of smart contracts \cite{buterin2014next}.
Defining such textual \acp{DSL} is made easier by complete frameworks and tools such as Xtext \cite{bettini2016implementing}. 

Graphical approaches to behavior specification are well represented by workflow and process languages.
\textit{Yet Another Workflow Language} (YAWL) \cite{van2005yawl} is a good example of a workflow language while \ac{BPMN} is the standard \cite{omg2011bpmn} for describing processes.
\ac{BPEL} \cite{weerawarana2005web} is another language used to orchestrate connections between different services, in the form of \ac{BPEL} workflows.
All such approaches target the need of describing execution sequences, complete with conditional transitions and extensions for integrating external services.
This is similar to how our approach defines behavioral flows, however the critical distinction is that in our approach the flows are defined using domain-specific elements previously created in a domain definition.
BPEL, YAWL or BPMN remain generic with respect to the application domain.
There are attempts to bring more semantics in BPMN-like languages.
For instance, in \cite{rodriguez2007bpmn} the authors propose an extension to model security requirements and in \cite{Brambilla:2012:CSW:2187980.2188014} the authors propose an extension for integrating \ac{BPMN} with social networking applications.
In \cite{Bocciarelli:2011:BEM:2048476.2048497} the authors propose an extension to model non-functional aspects, such as performance.
A comprehensive overview of various extensions brought by different works to the BPMN standard can be found in \cite{bpmexclass}.
Despite these attempts, such languages remain relatively generic and, crucially, do not bear domain semantics.
Each time a new process is defined in BPMN, its activities need to be configured in a technical environment to actually do something useful (some limited reuse exists in some tools, but this doesn't change the nature of the approach).
In contrast the flows that define the reusable functions in our approach are composed of already executable actions whose types are previously defined in a domain definition, and fully connected to the elements that make them ready to be executed.
Even more importantly, our approach provides the coordination mechanism that enables a clear distinction between what is executing and its representation on the client.
The clients in our approach mainly just need to know how to manage the generic data received from the cloud coordinator.
This is in contrast with how BPMN requires clients to behave.
Such clients are either fully integrated in the \acp{BPMS}, or, if they are separate applications, they fully depend on the process logic being executed. 

Related to the second axis, there are many approaches that target cloud support for application development.
Amazon Lambda \cite{amazonLambda} enables users to focus on writing code without the need to technically manage servers and containers.
IBM Cloud Foundry \cite{IBMCloudFoundry} also brings support for cloud-native applications in the form of fast deployment and management of user code.
In fact, the last few years saw rapid growth in the cloud-native space, with many initiatives \cite{IBMCloudFoundry2} focusing on facilitating access to containers, micro-services and infrastructure in ways that accelerate application development and deployment.
In relation to such work and initiatives, our execution approach sits at a higher level of abstraction but can certainly leverage their support.
Whereas cloud-native approaches typically involve support for technical programming languages, our engine directly executes domain-specific activities, and can decide how to break them down in individual tasks based on domain knowledge, on overall policies, on privacy and security settings and other considerations.
This is because the specification effectively understands what it executes.
It can, therefore, bring a level of optimization and smart reasoning to the actual execution.
This is impossible in engines that execute generic languages because such engines run generic tasks without ways to take run-time decisions based on domain semantics.
This is a core difference with of our approach where the execution ultimately adapts to the domain.

Related to the third axis, there exist efforts to develop generic GUI models, which can later be interpreted by device-independent applications, or generic applications \cite{Chmielewski2016DeclarativeGD}.
Other approaches aim at modeling generic client-server interfaces, which enable independent, generic  clients capable of interpreting the conversations with the cloud \cite{10.1145/2976767.2976776}.
However these approaches are highly technical and lack the ability to be easily extended and customized by users with little technical knowledge.
Furthermore, although generic, they only target screen-based devices, and as such they cannot support other common devices such as voice-based ones.

In novel areas where programming is simplified for end users, a popular approach is the use of visual programming (VP) systems \cite{educsci9030181, slany2012catroid}.
For example, to learn Internet of Things programming, Besari et. al \cite{8228460} allow users to connect different widgets which can be set with different characteristics in visually programmed Android applications.
In the domain of robot programming, CAPRICI provides a drag and drop interface that simulates the mental models of different people who work with robots \cite{beschi2019capirci}.
VP environments go beyond the textual forms of high level programming languages and express programming in graphical illustrative forms. They can either be general purpose or domain specific \cite{10.1145/169059.169119}.
Similarly to BPMN, general purpose VP systems use blocks and arrows to signify logic while domain specific VP systems use shapes meaningful to domain experts \cite{10.1145/169059.169119}.  
Most existing VP environments tend to be user, platform or domain specific \cite{ray2017survey} but they don't offer a mechanism equivalent to our coordination system that would enable the full application life-cycle described in this paper.
\section{Conclusion and Future Work}
\label{sec:conclusions}

This paper proposes an approach for creating applications where the client is decoupled from the application logic which is encoded on the cloud.
It does this by creating domain-specific descriptions of behavior, expressed as flows, and make them available for integration into any application.
The primary method by which we achieve this decoupling, and the primary contribution of this paper, is in the \textit{coordination mechanism}.
To validate the approach and evaluate its efficacy, we created a complete prototype implementation.
We used this prototype to implement an application and compared with similar implementations, as well as assess whether or not novices would be able to modify the generated client.

Based on our evaluation we can say that this framework will allow domain-experts -- or other types of users -- with limited technical knowledge to customize the client components of a fully functioning app.
We expect, due to the small amount of code that is necessary to implement complex flows, that these same domain experts will be able to implement a large amount of fully functioning back-ends running on the cloud as well.
One aspect that we did not evaluate, but remains true of our implementation, is that we are able to support a variety of options for user interface creation, ranging from web-based clients -- such as the one done in our evaluation -- as well as mobile apps, or even voice-based personal assistants.

The coordination mechanism enables the decoupling between the application behavior and the user interaction.
This, combined with the interpreted nature of the execution allows on-the-fly changes to behavioral logic to be propagated instantly to applications.
We believe that this can significantly reduce the entry barriers to application development.
For instance, it can be imagined that a community of user interface specialists provide a multitude of ``empty'' client applications, ready to integrate with any executable application logic written as flow-based functions.
This would allow people with little to no coding background to provide their domain knowledge directly in the form of executable, reusable functions and pick-and-choose from existing such ``empty'' client apps.
This is similar to how Wordpress blog writers choose a visual theme for their blog \cite{WordPressThemesWeb}.
Such plug-and-play composition would result in fully functioning self-updating applications, effectively created from the synergy of domain experts and platform-specific user interface experts. 

In the future, we are planning on further building and evaluating the impact of lowering technical barriers to application development.
We are interested in what people will build for themselves, and are planning on deploying to groups that have historically not had much say in the direction of the technology that they use, this will include rural parts of America, as well as with underrepresented groups.

\bibliography{our-papers.bib}
\bibliographystyle{ACM-Reference-Format}

\end{document}